\documentclass[preprint,12pt]{elsarticle}
\journal{International Journal of Engineering Science}
\usepackage{graphics,epsfig}
\usepackage{graphicx}
\usepackage{float}
\usepackage{amssymb,amstext,amsmath}
\usepackage{mathtools,physics,verbatim}
\usepackage{booktabs}
\usepackage{natbib}
\setcitestyle{round,aysep={,},yysep={,},citesep={;}}
\begin{document}
\newcommand{\balpha}{\boldsymbol{\alpha}}
\newcommand{\bbeta}{\boldsymbol{\beta}}
\newcommand{\bgamma}{\boldsymbol{\gamma}}
\newcommand{\bdelta}{\boldsymbol{\delta}}
\newcommand{\bepsilon}{\boldsymbol{\epsilon}}
\newcommand{\bvarepsilon}{\boldsymbol{\varepsilon}}
\newcommand{\bzeta}{\boldsymbol{\zeta}}
\newcommand{\bfoldeta}{\boldsymbol{\eta}}
\newcommand{\btheta}{\boldsymbol{\theta}}
\newcommand{\bvartheta}{\boldsymbol{\vartheta}}
\newcommand{\biota}{\boldsymbol{\iota}}
\newcommand{\bkappa}{\boldsymbol{\kappa}}
\newcommand{\blambda}{\boldsymbol{\lambda}}
\newcommand{\bmu}{\boldsymbol{\mu}}
\newcommand{\bnu}{\boldsymbol{\nu}}
\newcommand{\bxi}{\boldsymbol{\xi}}
\newcommand{\bpi}{\boldsymbol{\pi}}
\newcommand{\bvarpi}{\boldsymbol{\varpi}}
\newcommand{\brho}{\boldsymbol{\rho}}
\newcommand{\bvarrho}{\boldsymbol{\varrho}}
\newcommand{\bsigma}{\boldsymbol{\sigma}}
\newcommand{\bvarsigma}{\boldsymbol{\varsigma}}
\newcommand{\btau}{\boldsymbol{\tau}}
\newcommand{\bupsilon}{\boldsymbol{\upsilon}}
\newcommand{\bphi}{\boldsymbol{\phi}}
\newcommand{\bvarphi}{\boldsymbol{\varphi}}
\newcommand{\bchi}{\boldsymbol{\chi}}
\newcommand{\bpsi}{\boldsymbol{\psi}}
\newcommand{\bomega}{\boldsymbol{\omega}}
\newcommand{\bGamma}{\boldsymbol{\Gamma}}
\newcommand{\bDelta}{\boldsymbol{\Delta}}
\newcommand{\bTheta}{\boldsymbol{\Theta}}
\newcommand{\bLambda}{\boldsymbol{\Lambda}}
\newcommand{\bXi}{\boldsymbol{\Xi}}
\newcommand{\bPi}{\boldsymbol{\Pi}}
\newcommand{\bSigma}{\boldsymbol{\Sigma}}
\newcommand{\bUpsilon}{\boldsymbol{\Upsilon}}
\newcommand{\bPhi}{\boldsymbol{\Phi}}
\newcommand{\bPsi}{\boldsymbol{\Psi}}
\newcommand{\bOmega}{\boldsymbol{\Omega}}
\newcommand{\ldbracket}{[\![}
\newcommand{\rdbracket}{]\!]}
\newcommand{\ldangle}{\langle \!\langle}
\newcommand{\rdangle}{\rangle\!\rangle}
\def\Xint#1{\mathchoice
   {\XXint\displaystyle\textstyle{#1}}%
   {\XXint\textstyle\scriptstyle{#1}}%
   {\XXint\scriptstyle\scriptscriptstyle{#1}}%
   {\XXint\scriptscriptstyle\scriptscriptstyle{#1}}%
   \!\int}
\def\XXint#1#2#3{{\setbox0=\hbox{$#1{#2#3}{\int}$}
     \vcenter{\hbox{$#2#3$}}\kern-.5\wd0}}
\def\ddashint{\Xint=}
\def\dashint{\Xint-}
\newcommand{\operator}[1]{\mathop{\vphantom{\sum}\mathchoice
{\vcenter{\hbox{\huge $#1$}}}
{\vcenter{\hbox{\Large $#1$}}}{#1}{#1}}\displaylimits}
\newcommand{\opA}{\operator{\mathrm{A}}}

\begin{frontmatter}
\title{An asymptotically exact first-order shear deformation theory for functionally graded plates}
\author{K. C. Le$^{a,b}$\footnote{Phone: +84 93 4152458, email: lekhanhchau@tdtu.edu.vn}}
\address{$^a$Division of Computational Mathematics and Engineering, Institute for Computational Science, Ton Duc Thang University, Ho Chi Minh City, Vietnam
\\
$^b$Faculty of Civil Engineering, Ton Duc Thang University, Ho Chi Minh City, Vietnam}
\begin{abstract} 
An asymptotically exact first-order shear deformation theory for functionally graded elastic plates is derived using the variational-asymptotic method.  As an application, an analytical solution to the problem of wave propagation in a sandwich plate is found in accordance with this refined theory. Comparison between the dispersion curves obtained by 2-D plate theory and 3-D elasticity theory reveals that the former is accurate up to the order of $h^2/l^2$, where $h$ is the plate thickness and $l$ the wavelength. 
\end{abstract}


\end{frontmatter}

\section{Introduction}

\citet{reissner1945the} was the first to establish a refined plate theory accounting for transverse shear that bears his name and is now called the first-order shear deformation plate theory (FSDT).\footnote{One should also acknowledge an important contribution to this topic by \citet{mindlin1951influence}, although his motivation for including transverse shear and rotatory inertia was somewhat different: he wanted to capture the high-frequency thickness vibration of the plates.} He derived his theory from Castigliano's variational principle by making several hypotheses about the distribution of stresses on the transverse coordinate and by determining the energy in terms of stress resultants, bending moments, and shear forces. His pioneering work led to a large number of subsequent publications, mainly because of the applicability of his theory to moderately thick plates and the development of numerical methods \citep{bath1986a,arnold1989a,Bletzinger2000a,nguyen2010an,nguyen2017a}, but also because of the logic behind his derivation, which can be applied to other problems \citep{wang2000shear,batista2010an,challamel2019a}. There have been a number of attempts to justify or re-derive Reissner's plate theory based on Kantorovic-Krylov's semi-discrete method for the approximate solution of variational problems (see, for example, the review by \citet{wang2000shear}). However, the real advance in understanding the asymptotic nature of the Reissner derivation was made by \citet{berdichevsky1979variational,berdichevsky1979variational1,berdichevsky2009variational}. Using the variational-asymptotic method developed by himself, he gave the asymptotically exact derivation of the refined plate and shell theories accounting for transverse shear, which reduces to Reissner's theory for plates. He has shown that the integral characteristics of Reissner's theory are asymptotically accurate, but the distribution of displacements over the thickness are not.

Considering the wide application of functionally graded (FG) materials and structures, there is a large body of research proposing FSDT and higher order shear deformation theories (HSDT) for FG-plates \citep{reddy2000analysis,nguyen2008first,thai2013a,shen2016functionally,daikh2020,hirane2021}, among others\footnote{The literature on this topic is huge due to the variety of the 2-D theories for FG-plates: single-layer, multi-layer, refined theories including rotatory inertias and transverse shears et cetera. It is therefore impossible to cite all references. For the overview the reader may consult \citep{thai2015a,shen2016functionally} and the references therein.}. All the references cited above use the Kantorovic-Krylov's semi-discrete method, based on some assumptions about the kinematics of the deformation or a stress state through the thickness of the plates, to proceed with dimension reduction, so none of the obtained 2-D theories guarantees the asymptotic accuracy. The dynamic problem of wave propagation in FG-waveguides has been analyzed within the 3-D elasticity theory by \citet{lefebvre2001acoustic,baron2010propagation,kuznetsov2019abnormal,kuznetsov2021}. Since this analysis is not limited to low frequencies or long waves, its advantage is that all dispersion curves in the entire domain can be studied, yielding limits for low frequencies, points with zero group velocity, and zones with negative group velocity. The disadvantage of this 3-D approach is that the boundary conditions at the side edge of the finite plate lead to an infinite number of coupled equations for the amplitudes of modes of vibrations, which cannot be satisfied exactly. As for the asymptotic approaches, unfortunately, the author of this article is not currently aware of any asymptotically accurate FSDT for FG-plates. \citet{berdichevsky2009variational} has tried to construct the refined theory for anisotropic and inhomogeneous shells, which takes into account the transverse shear under a constraint for the elastic moduli, but the constraint of the constant Poisson ratio is too restrictive, so almost all real FG-plates do not satisfy it. For laminated plates, which are a special case of FG-plates, an asymptotic analysis of the energy functional was provided by \citet{sutyrin1997derivation} and \citet{yu2005mathematical}. However, since in the general case of the laminated plates the dimension reduction does not lead to an FSDT, they have tried to optimize the parameters so that a derived theory is as close as possible to asymptotic correctness while being an FSDT. It is also worth mentioning some recent applications of the variational-asymptotic method to plates and shells and to homogenization in \citep{le2013,le2016asymptotically,le2017,le2020,le2020a,shi2021,phanendra2022}.

The objective of this paper is to construct the asymptotically exact FSDT for FG-plates by means of the variational-asymptotic method. Our goal is to construct the refined plate theory with the asymptotic accuracy up to the order of $h^2/l^2$, where $h$ is the plate thickness and $l$ is the wavelength. Since we consider only the inhomogeneity in the transverse direction, we assume that the mass density and elastic moduli of the isotropic elastic FG-plate vary in this direction such that their distributions are even functions of the transverse coordinate. The dimension reduction is based on the asymptotic analysis of the action functional containing a small parameter $h/l$, which allows to obtain the distribution of displacements from the solution of the thickness problem. The obtained average energy must also be extrapolated to ensure the correct behavior at short waves. As a result of this asymptotic analysis, the FSDT is obtained, where all the coefficients of the theory are given in closed analytical form and can be calculated for any FG-plate satisfying the above assumption. We apply this theory to the problem of wave propagation in a sandwich plate and compare the dispersion curves resulting from the proposed theory and the elasticity theory. We show that the former is exact up to the order of $h^2/l^2$, which justifies our asymptotic analysis.

The paper is organized as follows. After this brief introduction, the variational formulation of the problem is given in Section 2. Section 3 is devoted to the asymptotic analysis of the action functional. Section 4 presents the two-dimensional FSDT for FG-plates. Section 5 analyzes the wave propagation in a sandwich plate and shows the comparison of the dispersion curves. Finally, Section 6 concludes the paper.

\section{Variational formulation for FG-plates}
\begin{figure}[htb]
	\centering
	\includegraphics[width=8cm]{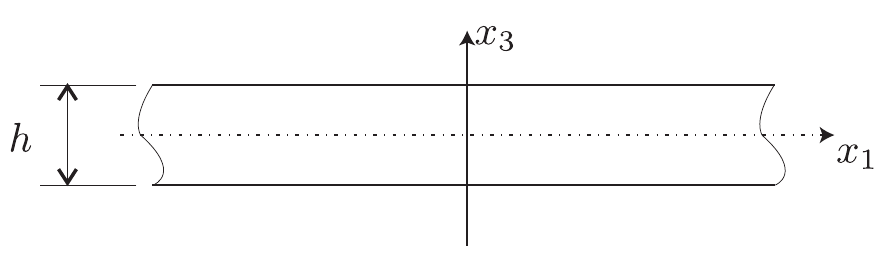}
	\caption{Cross section of a plate}
	\label{fig:1}
\end{figure}

Let $\Omega $ be a two-dimensional domain in the $(x_1,x_2)$-plane bounded by a smooth closed curve $\partial \Omega $. We consider a plate which in the undeformed state occupies the 3-D region $\mathcal{V}=\Omega \times (-h/2,h/2)$. Its cross section in the plane $(x_1,x_3)$ is shown in Fig.~\ref{fig:1}. We call $\Omega $ the plate mid-plane and $h$ its thickness. 

We analyze the forced vibration of the plate made of a functionally graded isotropic elastic material whose mass density and elastic moduli depend on $x_3$. The action functional reads 
\begin{multline}\label{eq:1}
I[\vb{w}(\vb{x},t)]=\int_{t_0}^{t_1}\int_{\Omega}\int_{-h/2}^{h/2}[T(x_3,\dot{\vb{w}})-W(x_3,\vb*{\varepsilon})] \dd{x_3}\dd{a}\dd{t}
\\
+\int_{t_0}^{t_1}\int_{\Omega} (\vb*{\tau }\vdot \vb{w} |_{x_3=h/2} +\vb*{\tau }\vdot \vb{w}  |_{x_3=-h/2})\dd{a}\dd{t},
\end{multline}
where $\dd a=\dd x_1\dd x_2$ denotes the area element of the mid-plane, $\vb{w}(\vb{x},t)$ is the displacement field, while $\vb*{\tau }$ is the traction acting on the upper and lower faces of the plate. The explicit dependence of the Lagrangian on the transverse coordinate $x_3$ of this functionally graded material is precisely indicated. The kinetic energy density equals 
\begin{equation*}
T(x_3,\dot{\vb{w}})=\frac{1}{2}\rho (x_3) \dot{\vb{w}}\vdot \dot{\vb{w}}=\frac{1}{2}\rho (x_3) (\dot{w}_\alpha \dot{w}_\alpha +\dot{w}^2),
\end{equation*}
with dot denoting the time derivative. The stored energy density $W$ reads
\begin{equation*}
W(x_3,\vb*{\varepsilon})=\frac{1}{2}\lambda(x_3)(\varepsilon _{ii})^2 +\mu(x_3)\varepsilon _{ij}\varepsilon _{ij},
\end{equation*}
with the components of the strain tensor $\vb*{\varepsilon}$ being given by
\begin{displaymath}
\varepsilon_{ij}=\frac{1}{2}(w_{i,j}+w_{j,i})\equiv w_{(i,j)}.
\end{displaymath}
We additionally assume that $\rho (x_3)$, $\lambda (x_3)$, and $\mu (x_3)$ are even functions of $x_3$. Here and in the following, we use Latin indices, running from 1 to 3, to refer to the spatial co-ordinates and the Greek indices, running from 1 to 2, to refer to the plane co-ordinates $x_1$ and $x_2$. The comma before an index denotes differentiation with respect to the corresponding coordinate, the parentheses surrounding a pair of indices denote the symmetrization operation, and summation over repeated indices is understood.

Hamilton's variational principle states that the true displacement $\check{\mathbf{w}}(\mathbf{x},t)$ of a functionally graded plate change in space and time in such a way that the action functional \eqref{eq:1} becomes stationary at $\check{\mathbf{w}}(\mathbf{x},t)$ among all continuously differentiable functions $\mathbf{w}(\mathbf{x},t)$ satisfying the initial and end conditions as well as the kinematic boundary conditions (see, e.g., \citep{berdichevsky2009variational}). The problem is to replace the three-dimensional action functional by an approximate two-dimensional action functional for a thin FG-plate, whose functions depend only on the longitudinal co-ordinates $x_1,x_2$ and time $t$. The possibility of dimension reduction is related to the smallness of the ratio between the thickness $h$ and the characteristic scale of change of the deformation state in the longitudinal directions, $l$, (see \citep{le1999vibrations} and Section 3). We assume that
\begin{displaymath}
\frac{h}{l}\ll 1.
\end{displaymath}
Additionally, we assume that 
\begin{equation}\label{eq:2}
\frac{h}{ c\tau }\ll 1,
\end{equation}
where $\tau $ is the characteristic scale of change of the function $\check{\mathbf{w}}(\mathbf{x},t)$ in time (see \citep{le1999vibrations}) and $c$ the minimal velocity of plane waves in the elastic material under consideration. This means that we consider only statics or low-frequency vibrations of the functionally graded elastic plate. By using the variational-asymptotic method, the two-dimensional action functional will be constructed below in which terms up to the order $h^2/l^2$ are exact compared to unity (the refined plate theory accounting for transverse shear, now called FSDT).

In order to fix the domain of the transverse co-ordinate in the passage to the limit $h\to 0$, we introduce the dimensionless co-ordinate
\begin{equation*}
\zeta =\frac{x_3}{h}, \quad \zeta \in [-1/2,1/2],
\end{equation*}
and transform the action functional to
\begin{multline}\label{eq:3}
I[\vb{w}(\vb{x},t)]=\int_{t_0}^{t_1}\int_{\Omega}\int_{-1/2}^{1/2}h[T(\zeta ,\dot{\mathbf{w}})-W(\zeta ,\vb*{\varepsilon})] \dd{\zeta }\dd{a}\dd{t}
\\
+\int_{t_0}^{t_1}\int_{\Omega} (\tau_i w_i |_{\zeta=1/2} +\tau_i w_i |_{\zeta=-1/2})\dd{a}\dd{t}.
\end{multline}
Now $h$ enters the action functional explicitly through the components of the strain tensor $\varepsilon _{ij}$ 
\begin{equation*}
\varepsilon _{\alpha \beta }=
w_{(\alpha ,\beta )},
\quad 
2\varepsilon _{\alpha 3}= 
\frac{1}{h}w_{\alpha |\zeta }+w_{,\alpha }, \quad \varepsilon _{33}=\frac{1}{h}w_{|\zeta }.
\end{equation*}
The vertical bar followed by $\zeta $ indicates the partial derivative with respect to $\zeta $ and {\it not} with respect to $x_\zeta $. 

\section{\bf Asymptotic analysis of the action functional}

We restrict ourselves to the low frequency vibrations of the FG-plate for which assumption \eqref{eq:2} is valid. Based on this assumption we may neglect the kinetic energy density in the variational-asymptotic procedure of finding the asymptotic expansion of the displacement field.\footnote{For the high-frequency vibrations of elastic shells and rods where the kinetic energy density should be kept in the variational-asymptotic analysis see \citep{berdichevsky1980high,berdichevsky1982high,le1997high,le1999vibrations,kaplunov1998}.}  Likewise, the work of external traction $\vb*{\tau}$ can also be neglected during this procedure. The average kinetic energy density and the work of external traction can then be calculated once the asymptotic formulas for the displacements are found. Before applying the variational-asymptotic procedure to functional \eqref{eq:3} let us transform the stored energy density to another form more convenient for the asymptotic analysis. We note that, among terms of $W(\varepsilon _{ij})$, the derivatives $w_{\alpha |\zeta }/h$ and $w_{|\zeta }/h$ in $\varepsilon _{\alpha 3}$ and $\varepsilon _{33}$ are the main ones in the asymptotic sense. Therefore it is convenient to single out the components $\varepsilon _{\alpha 3}$ and $\varepsilon _{33}$ in the stored energy density. To this end we represent the latter as the sum of two quadratic forms $W_\parallel $ and $W_\perp $ corresponding to longitudinal and transverse stored energy densities, respectively. These are defined by
\begin{equation}
W_\parallel =\min_{\varepsilon _{\alpha 3},
\varepsilon _{33}}W, \quad
W_\perp =W-W_\parallel . \label{eq:4}
\end{equation}
Simple calculations give
\begin{align*}
W_\parallel &=\mu [\sigma (\varepsilon _{\alpha \alpha})^2 +\varepsilon _{\alpha \beta }\varepsilon _{\alpha \beta }],
\\ 
W_\perp &=\frac{1}{2}(\lambda +2\mu )(\epsilon _{33}+ 
\sigma \varepsilon _{\alpha \alpha})^2
+2\mu \varepsilon _{\alpha 3}\varepsilon _{\alpha 3},
\end{align*}
where $\sigma =\lambda /(\lambda +2\mu )=\nu/(1-\nu)$. For FG-plate $\lambda$, $\mu$, and $\sigma$ are functions of $\zeta$. The $\zeta$-argument is omitted briefly in these functions when a precise specification is not required.

We could start the variational-asymptotic procedure with the determination of the set $\mathcal{N}$ according to its general scheme \citep{le1999vibrations,berdichevsky2009variational}. As a result, it would turn out that, at the first step, the function ${\bf w}$ does not depend on the transverse co-ordinate $\zeta $: ${\bf w}={\bf u}(x_\alpha ,t)$; at the second step additional degrees of freedom $\varphi_\alpha(x_\alpha,t)$ associated with transverse shear occur in ${\bf w}^* $; and at the next step ${\bf w}^{**}$ is completely determined through ${\bf u}$ and $\varphi _\alpha$. Thus, the set $\mathcal{N}$ according to the variational-asymptotic scheme consists of functions ${\bf u}(x_\alpha ,t)$ and $\varphi _\alpha(x_\alpha ,t)$. We will pass over these long, but otherwise standard, derivations and make the following change of unknown functions 
\begin{equation}\label{eq:5} 
\begin{split}
&w_\alpha (x_\alpha ,\zeta ,t)=u_\alpha (x_\alpha ,t)+h\varphi _{\alpha } (x_\alpha ,t)\zeta -hu_{,\alpha } \zeta +hy_\alpha (x_\alpha ,\zeta ,t),
\\
&w(x_\alpha ,\zeta ,t)=u(x_\alpha ,t)-h\mathcal{I}[\sigma]u_{\beta ,\beta}+hy (x_\alpha ,\zeta ,t).
\end{split}
\end{equation}
In these formulas $u_\alpha , u$ correspond to the mean displacements of the plate, while $\psi _\alpha=\varphi_\alpha -u_{,\alpha}$ describe the mean rotation angles\footnote{It is easy to see that $\varphi_\alpha$ are the rotation angles due to the transverse shear. If $\varphi_\alpha=0$, the shear strains $\varepsilon_{\alpha 3}$ vanish in the first approximation, and the transverse fibers remain ``on average'' perpendicular to the deformed mid-plane as in the Kirchhoff's plate theory.} in the following sense
\begin{equation}\label{eq:6}
\begin{split}
&u_\alpha (x_\alpha ,t)=\langle w_\alpha (x_\alpha ,\zeta ,t)\rangle ,\quad u(x_\alpha ,t)=\langle w(x_\alpha ,\zeta ,t) \rangle ,
\\ 
&\psi _\alpha(x_\alpha ,t)=\langle w_\alpha (x_\alpha ,\zeta ,t)\zeta \rangle /(h/12),
\end{split}
\end{equation}
where $\langle . \rangle $ denotes the integration over $\zeta $ within the limits $[-1/2,1/2]$. By $\mathcal{I}[f]$ we mean the following transformation applied to an arbitrary function $f(\zeta)$
\begin{displaymath}
\mathcal{I}[f](\zeta)=\int_0^\zeta f(\xi)\dd{\xi}-\langle \int_0^\zeta f(\xi)\dd{\xi} \rangle .
\end{displaymath}
By this definition, $\mathcal{I}[f](\zeta)$ is an odd function if $f(\zeta)$ is an even function and vice versa. Moreover, $\langle \mathcal{I}[f] \rangle =0$. 

Because of definitions \eqref{eq:6} functions $y_\alpha $ and $y$ should satisfy the following constraints
\begin{equation}
\langle y_\alpha \rangle =0,\quad \langle y_\alpha \zeta \rangle =0,\quad \langle y \rangle =0. \label{eq:7}
\end{equation}
Equations \eqref{eq:5}-\eqref{eq:7} set up a one-to-one correspondence between $w_\alpha , w$ and the set of functions $u_\alpha ,u, \varphi_\alpha ,y_\alpha , y$ and determine the change in the unknown functions $\{ w_\alpha ,w\} \to \{
u_\alpha ,u, \varphi_\alpha ,y_\alpha , y\}$.

Asymptotic analysis enables one to determine the order of smallness of $y_\alpha ,y$. If these terms are neglected, then the deformation state of a plate is characterized solely by the measures of extension $A_{\alpha \beta }=u_{(\alpha ,\beta )} $ and the measures of bending $B_{\alpha \beta }=u_{,\alpha \beta }
-\varphi _{(\alpha ,\beta )}$. We introduce the following notation
\begin{gather*}
\varepsilon _A=\max_{\Omega }\sqrt{A_{\alpha \beta 
}A^{\alpha \beta }}, \quad \varepsilon _B=h\max_{\Omega }
\sqrt{B_{\alpha \beta }B^{\alpha \beta }},
\\
\Delta _\alpha =\max_{\mathcal{V}}|y_{\alpha |\zeta }|,\quad
\Delta =\max_{\mathcal{V}}|y_{|\zeta }| .
\end{gather*}
Consider a certain point of the mid-plane $\Omega $. The best constant $l$ in the inequalities
\begin{equation*}
\begin{split}
\left| A_{\alpha \beta ,\gamma }\right| \le \frac{\varepsilon _A}{l}, \quad h\left| B_{\alpha \beta ,\gamma }\right| 
\le \frac{\varepsilon _B}{l}, \quad \left| \varphi_{\alpha}\right| \le \frac{h}{l}(\varepsilon _A+\varepsilon _B)
\\
\left| \varphi_{\alpha,\beta }\right| \le \frac{h}{l^2}(\varepsilon _A+\varepsilon _B),\quad \max_\zeta \left| y_{\alpha ,\beta }\right| \le \frac{\Delta _\alpha}{l} 
,  \quad \max_\zeta \left| y_{,\alpha }\right| 
\le \frac{\Delta }{l}
\end{split}
\end{equation*}
is called the characteristic scale of change of the deformation state in the longitudinal directions. We define the inner domain $\Omega _i$ as a subdomain of $\Omega$ in which the following inequality holds:
\begin{equation*}
h/l\ll 1.
\end{equation*}
We assume the domain $\Omega $ to consist of the inner domain $\Omega _i$ and a domain $\Omega _b$ abutting on the contour $\partial \Omega$ with width of the order $h$ (boundary layer). Then functional \eqref{eq:3} can be decomposed into the sum of two functionals, an inner one for which an iteration process will be applied, and a boundary layer functional. We concentrate first on the inner functional.

We now fix $u_\alpha , u, \varphi_\alpha $ and seek $y_\alpha , y$. Substituting \eqref{eq:5} into the action functional \eqref{eq:3}, we will keep in it the asymptotically principal terms containing $y_\alpha , y$ and neglect all other terms. The estimations based on the above inequalities lead to the asymptotic formulas
\begin{equation}
\begin{split}
&\varepsilon _{33}=-\sigma A_{\beta \beta}+y_{|\zeta }, \quad 2\varepsilon _{\alpha 3}=\varphi_\alpha +y_{\alpha |\zeta }-h\mathcal{I}[\sigma]A_{\beta \beta,\alpha}+hy_{,\alpha}, 
 \\
&\varepsilon _{\alpha \beta }=A_{\alpha \beta }- hB_{\alpha \beta }\zeta +hy_{(\alpha ,\beta)}.
\label{eq:8}
\end{split}
\end{equation}
According to formulas \eqref{eq:8} the longitudinal stored energy density does not contain asymptotically principal terms containing $y_\alpha , y$ and can be neglected. Since the asymptotically principal terms in the transverse stored energy density contains only the derivatives of $y_\alpha ,y $ with respect to $\zeta $, we drop the integration over $\Omega $ and $t$ and reduce the thickness problem to finding minimum of the functional
\begin{multline*}
I_\perp =\frac{h}{2} \langle 
[ (\lambda+2\mu)(y_{|\zeta }-\sigma hB_{\beta \beta }\zeta)^2 
\\
+\mu (\varphi_\alpha +y_{\alpha |\zeta }-h\mathcal{I}[\sigma]A_{\beta \beta,\alpha}+hy_{,\alpha})(\varphi_\alpha +y_{\alpha |\zeta }-h\mathcal{I}[\sigma]A_{\gamma \gamma,\alpha}+hy_{,\alpha}) ]\rangle .
\end{multline*}
This problem can be solved in two steps. We first minimize the functional
\begin{displaymath}
\frac{h}{2} \langle 
(\lambda+2\mu)(y_{|\zeta }-\sigma hB_{\beta \beta }\zeta)^2\rangle 
\end{displaymath}
under constraint \eqref{eq:7}$_3$. Obviously, the minimum is equal to zero and is attained at
\begin{equation}\label{eq:9}
y=h\mathcal{I}[\sigma \zeta]B_{\beta \beta}.
\end{equation}

Now, substituting $y$ from here into the remaining term of $I_\perp$, we arrive at the second step: Minimize
\begin{displaymath}
\frac{h}{2}\langle \mu (\varphi_\alpha +y_{\alpha |\zeta }-h\mathcal{I}[\sigma]A_{\beta \beta,\alpha}+h^2\mathcal{I}[\sigma \zeta]B_{\beta \beta,\alpha })(\varphi_\alpha +y_{\alpha |\zeta }-h\mathcal{I}[\sigma]A_{\gamma \gamma,\alpha}+h^2\mathcal{I}[\sigma \zeta]B_{\gamma \gamma,\alpha }) \rangle 
\end{displaymath}
among $y_\alpha $ satisfying the first two constraints in \eqref{eq:7}. To simplify this minimization problem we change the unknown functions
\begin{equation}\label{eq:10}
y_\alpha = f(\zeta)hA_{\beta \beta,\alpha}+g(\zeta)h^2B_{\beta \beta,\alpha}+z_\alpha
\end{equation}
such that constraints \eqref{eq:7}$_{1,2}$ are fulfilled and
\begin{equation*}
y_{\alpha |\zeta }=h\mathcal{I}[\sigma]A_{\beta \beta,\alpha}-h^2\mathcal{I}[\sigma \zeta]B_{\beta \beta,\alpha }+\alpha h^2B_{\beta \beta,\alpha }+z_{\alpha |\zeta },
\end{equation*}
where $\alpha$ is still an unknown coefficient. Then
\begin{equation*}
f(\zeta)=\mathcal{I}[\mathcal{I}[\sigma]], \quad g(\zeta)=-\mathcal{I}[\mathcal{I}[\sigma \zeta]]+\alpha \zeta .
\end{equation*}
Since $\sigma(\zeta)$ is even function, $f(\zeta)$ is also even function and $\langle f\rangle =0$. Therefore the first term on the right-hand side of \eqref{eq:10} satisfies the constraints \eqref{eq:7}$_{1,2}$. Function $g(\zeta)$ is odd function, so the constraint \eqref{eq:7}$_{1}$ is automatically satisfied. To fulfill the second constraint $\langle g(\zeta)\zeta \rangle =0$ the coefficient $\alpha$ must be
\begin{equation}
\label{eq:11}
\alpha=12\langle \zeta \mathcal{I}[\mathcal{I}[\sigma \zeta]]\rangle .
\end{equation}
With \eqref{eq:10} we reduce the problem to minimizing the following functional
\begin{equation*}
J_\perp=\frac{h}{2}\langle \mu (z_{\alpha |\zeta }+\bar{\varphi}_\alpha )(z_{\alpha |\zeta }+\bar{\varphi}_\alpha )\rangle 
\end{equation*}
among $z_\alpha$ satisfying the constraints $\langle z_{\alpha }\rangle =\langle z_{\alpha }\zeta \rangle =0$, where 
\begin{equation}\label{eq:12}
\bar{\varphi}_\alpha =\varphi_\alpha +\alpha h^2B_{\beta \beta,\alpha }.
\end{equation}

To solve this minimization problem we consider the dual variational problem \citep{berdichevsky2009variational}
\begin{equation*}
\min J_\perp=\max J_\perp^*, \quad J_\perp^*=h \langle p_\alpha \bar{\varphi}_\alpha -\frac{1}{2\mu }p_\alpha p_\alpha \rangle .
\end{equation*}
The maximum is sought among $p_\alpha$ and the numbers $a_\alpha$ and $b_\alpha$ such that
\begin{displaymath}
p_{\alpha |\zeta }=a_\alpha+b_\alpha \zeta ,\quad p_\alpha (\pm 1/2)=0.
\end{displaymath}
The numbers $a_\alpha$ and $b_\alpha $ are the Lagrange multipliers for the constraints $\langle z_{\alpha }\rangle =0$ and $\langle z_{\alpha }\zeta \rangle =0$. Integrating the equations for $p_\alpha$ and using the boundary conditions for them, we find that $a_\alpha=0$ and $p_\alpha =b_\alpha k(\zeta)$, where $k(\zeta)=\frac{1}{2}(\zeta^2-1/4)$. Substituting $p_\alpha$ into $J_\perp^*$ and maximizing with respect to $b_\alpha$, we get
\begin{equation*}
\Phi_\perp=\min J_\perp=\max J_\perp^*=\frac{1}{2}h\mu ^* \bar{\varphi}_\alpha \bar{\varphi}_\alpha , \quad \mu^*=\frac{\langle k\rangle ^2}{\langle k^2/\mu \rangle }.
\end{equation*}
Since the $b_\alpha$ that give maximum to $J_\perp^*$ are
\begin{displaymath}
b_\alpha=\frac{\langle k\rangle }{\langle k^2/\mu \rangle }\bar{\varphi}_\alpha,
\end{displaymath}
we have
\begin{displaymath}
p_\alpha =b_\alpha k(\zeta)=\frac{\langle k\rangle k}{\langle k^2/\mu \rangle }\bar{\varphi}_\alpha .
\end{displaymath}
Using the relation between minimizer and maximizer
\begin{displaymath}
z_{\alpha |\zeta }+\bar{\varphi}_\alpha =\frac{1}{\mu }p_\alpha ,
\end{displaymath}
we find also $z_\alpha $ from here
\begin{equation}
\label{eq:13}
z_\alpha =\mathcal{I} \Bigl[ \frac{\langle k\rangle k}{\langle k^2/\mu \rangle \mu}-1\Bigr] \bar{\varphi}_\alpha .
\end{equation}

We want now to compute the average longitudinal energy density. Taking into account \eqref{eq:10}, it is sufficient to approximate $\varepsilon_{\alpha \beta}$ in this refined theory by
\begin{displaymath}
\varepsilon _{\alpha \beta }=A_{\alpha \beta }- hB_{\alpha \beta }\zeta +\mathcal{I}[\mathcal{I}[\sigma]] h^2A_{\gamma \gamma, \alpha \beta }
\end{displaymath}
because other terms do not bring correction of order $h^2/l^2$. Substituting $\varepsilon_{\alpha \beta}$ into $W_\parallel$ and integrating over the thickness, we obtain
\begin{multline*}
\Phi_\parallel=h[\langle \mu \sigma\rangle (A_{\alpha \alpha})^2+\langle \mu \rangle A_{\alpha \beta}A_{\alpha \beta}+\langle \mu \sigma \zeta^2 \rangle h^2(B_{\alpha \alpha})^2+\langle \mu \zeta^2 \rangle h^2 B_{\alpha \beta}B_{\alpha \beta}
\\
+2\langle \mu \sigma \mathcal{I}[\mathcal{I}[\sigma]]\rangle h^2A_{\delta \delta} A_{\gamma \gamma, \alpha \alpha}+2\langle \mu \mathcal{I}[\mathcal{I}[\sigma]]\rangle h^2A_{\alpha \beta } A_{\gamma \gamma, \alpha \beta}].
\end{multline*}
Note that, due to the evenness of elastic moduli, there is no cross term between $A_{\alpha \beta}$ and $B_{\alpha \beta}$. However, in contrast to the refined theory for homogeneous plates, there are two additional cross terms between $A_{\alpha \beta}$ and $A_{\gamma \gamma, \alpha \beta}$ due to the dependence of $\mu$ and $\sigma$ on $\zeta$. Adding the transverse and longitudinal stored energy density together, we obtain the total average stored energy density in the form
\begin{multline}
\label{eq:14}
\Phi =h[\langle \mu \sigma\rangle (A_{\alpha \alpha})^2+\langle \mu \rangle A_{\alpha \beta}A_{\alpha \beta}+\langle \mu \sigma \zeta^2 \rangle h^2(B_{\alpha \alpha})^2+\langle \mu \zeta^2 \rangle h^2 B_{\alpha \beta}B_{\alpha \beta}
\\
+2\langle \mu \sigma \mathcal{I}[\mathcal{I}[\sigma]]\rangle h^2A_{\delta \delta} A_{\gamma \gamma, \alpha \alpha}+2\langle \mu \mathcal{I}[\mathcal{I}[\sigma]]\rangle h^2A_{\alpha \beta } A_{\gamma \gamma, \alpha \beta} +\frac{1}{2}\mu ^* \bar{\varphi }_\alpha \bar{\varphi}_{\alpha }] .
\end{multline}

We turn next to the average kinetic energy density. To find the latter we must first compute the velocity based on the asymptotic expansion \eqref{eq:5}. Within the desired accuracy of the refined theory \citep{berdichevsky2009variational} we can neglect $\dot{y}_\alpha$ and $\dot{y}$ as compared to other time derivatives, so
\begin{displaymath}
\dot{w}_\alpha =\dot{u}_\alpha +\dot{\psi}_\alpha \zeta ,\quad \dot{w}=\dot{u}-h\mathcal{I}[\sigma]\dot{u}_{\alpha ,\alpha }.
\end{displaymath}
Substituting these formulas into the kinetic energy density and integrating over the thickness, we find that
\begin{equation*}
\Theta=\frac{h}{2}[ \langle \rho \rangle \dot{u}_\alpha \dot{u}_\alpha +\langle \rho \rangle \dot{u}^2+\langle \rho \zeta^2 \rangle h^2 \dot{\psi}_\alpha \dot{\psi}_\alpha +\langle \rho (\mathcal{I}[\sigma])^2 \rangle h^2(\dot{u}_{\alpha ,\alpha })^2 ] .
\end{equation*}
Again, the evenness of $\rho$ and the oddness of $\zeta $ and $\mathcal{I}[\sigma]$ ensure that no cross term occurs between $\dot{u}_{,\alpha }$ and $\dot{\psi}_\alpha$ as well as between $\dot{u}$ and $\dot{u}_{\alpha ,\alpha }$. The average kinetic energy density turns out to be simple enough. 

Combining the average stored and kinetic energy densities, we obtain the average 2-D action functional in the form
\begin{equation*}
J[u_\alpha ,u,\psi _\alpha ]=\int_{t_0}^{t_1}\int_{\Omega }[\Theta (\dot{u}_\alpha, \dot{u},\dot{\psi }_\alpha)-\Phi (A_{\alpha \beta },B_{\alpha \beta },A_{\gamma \gamma,\alpha \beta},\bar{\varphi }_{\alpha })]\dd{a} \dd{t},
\end{equation*} 
with $\Theta$ and $\Phi$ given above. This action functional can still be simplified. First, if we introduce a new unknown function
\begin{equation}
\label{eq:15}
\bar{u}=u+\alpha h^2B_{\alpha \alpha}.
\end{equation}
then the rotation angles $\psi_\alpha =\varphi_\alpha -u_{,\alpha}$ do not alter their form after these changes of unknown functions \eqref{eq:12} and \eqref{eq:15},
\begin{displaymath}
\psi _\alpha =\varphi_\alpha -u_{,\alpha}=\bar{\varphi}_\alpha -\bar{u}_{,\alpha},
\end{displaymath}
while $B_{\alpha \beta}$, expressed in terms of $\psi_{\alpha }$, do not contain the second derivatives
\begin{equation*}
B_{\alpha \beta}=-\psi_{(\alpha ,\beta)}.
\end{equation*}
Therefore, it is convenient to choose $\psi_\alpha$ together with $u_\alpha$ and $\bar{u}$ as the primary degrees of freedom and express the total average stored energy density in their terms. However, by making the change of unknown functions $u\to \bar{u}$ according to \eqref{eq:15} $\Theta$ becomes after neglecting terms of higher order of smallness
\begin{displaymath}
\Theta=\frac{h}{2}[ \langle \rho \rangle \dot{u}_\alpha \dot{u}_\alpha +\langle \rho \rangle \dot{\bar{u}}^2-2\langle \rho \rangle \alpha h^2\dot{\bar{u}}\dot{\bar{u}}_{,\alpha \alpha}+\langle \rho \zeta^2 \rangle h^2 \dot{\psi}_\alpha \dot{\psi}_\alpha +\langle \rho (\mathcal{I}[\sigma])^2 \rangle h^2(\dot{u}_{\alpha ,\alpha })^2 ] .
\end{displaymath}
This average kinetic energy density is asymptotically accurate (up to the order $h^2/l^2$) for long waves but not suitable for short waves because it is not positive definite. If we integrate the third term by parts, neglecting the divergence term going to the boundary (as a null Lagrangian), and adding small terms involving $\dot{\bar{\varphi}}_\alpha$, we get the meaningful short wave extrapolation \citep{berdichevsky1979variational,le1999vibrations} of the previous formula
\begin{equation*}
\Theta=\frac{h}{2}[ \langle \rho \rangle \dot{u}_\alpha \dot{u}_\alpha +\langle \rho \rangle \dot{\bar{u}}^2+\langle \rho \zeta^2+2\rho \alpha \rangle h^2 \dot{\psi}_\alpha \dot{\psi}_\alpha +\langle \rho (\mathcal{I}[\sigma])^2 \rangle h^2 (\dot{u}_{\alpha ,\alpha })^2 ] .
\end{equation*} 
Second, there is another short-wave extrapolation where the fifth and sixth term in $\Phi_\parallel$, which are not positive definite and behave ``badly'' at short waves, can be transformed and combined with the last term in $\Theta$. Indeed, by integration by parts and neglecting the null-Lagrangian, we reduce these two terms to
\begin{equation}\label{eq:16}
2(\langle \mu \sigma \mathcal{I}[\mathcal{I}[\sigma]]\rangle +\langle \mu \mathcal{I}[\mathcal{I}[\sigma]]\rangle) h^2A_{\beta \beta}A_{\gamma \gamma,\alpha \alpha}. 
\end{equation}
By observing that, in the first approximation (see next Section)
\begin{displaymath}
\langle \rho \rangle \ddot{u}_\alpha =2\langle \mu \sigma \rangle A_{\gamma \gamma, \alpha} +2\langle \mu \rangle A_{\alpha \beta, \beta},
\end{displaymath}
so that
\begin{displaymath}
\langle \rho \rangle \ddot{u}_{\alpha ,\alpha }=2(\langle \mu \sigma \rangle +\langle \mu \rangle )A_{\gamma \gamma, \alpha \alpha},
\end{displaymath}
we can replace the last factor $A_{\gamma \gamma, \alpha \alpha}$ of the correction term in \eqref{eq:16} by 
$$\frac{\langle \rho \rangle }{2(\langle \mu \sigma \rangle +\langle \mu \rangle )}\ddot{u}_{\alpha ,\alpha }.$$
Then, doing the integration by parts in time and neglecting the null-Lagrangian, we can combine the last term in $\Theta$ with this term to $(h/2)\rho^* h^2(\dot{u}_{\alpha ,\alpha })^2$,
where
\begin{displaymath}
\rho^*=\langle \rho (\mathcal{I}[\sigma])^2 \rangle -\frac{2\langle \rho \rangle (\langle \mu \sigma \mathcal{I}[\mathcal{I}[\sigma]]\rangle +\langle \mu \mathcal{I}[\mathcal{I}[\sigma]]\rangle)}{\langle \mu \sigma \rangle +\langle \mu \rangle } .
\end{displaymath}
Thus, the final formulas for the average stored and kinetic energy density after this short-wave extrapolation are as follows 
\begin{multline}
\label{eq:17}
\Phi =h[\langle \mu \sigma\rangle (A_{\alpha \alpha})^2+\langle \mu \rangle A_{\alpha \beta}A_{\alpha \beta}+\langle \mu \sigma \zeta^2 \rangle h^2(B_{\alpha \alpha})^2+\langle \mu \zeta^2 \rangle h^2 B_{\alpha \beta}B_{\alpha \beta}
\\
+\frac{1}{2}\mu ^* (\psi _\alpha +\bar{u}_{,\alpha })(\psi _\alpha +\bar{u}_{,\alpha })] ,
\end{multline}
and
\begin{equation}
\label{eq:18}
\Theta=\frac{h}{2}[ \langle \rho \rangle \dot{u}_\alpha \dot{u}_\alpha +\langle \rho \rangle \dot{\bar{u}}^2+\langle \rho \zeta^2+2\rho \alpha \rangle h^2 \dot{\psi}_\alpha \dot{\psi}_\alpha +\rho^* h^2 (\dot{u}_{\alpha ,\alpha })^2 ] .
\end{equation} 

We turn finally to the work of external tractions $\tau_\alpha$ and $\tau$. Substituting $w_\alpha$ and $w$ from \eqref{eq:5} into the last integral of \eqref{eq:3} and taking into account Eqs.~\eqref{eq:9}, \eqref{eq:10}, \eqref{eq:11}, and \eqref{eq:13}, we find that
\begin{multline*}
\mathcal{A}=\int_{t_0}^{t_1}\int_{\Omega} \Bigl\{ P_\alpha u_\alpha +P u +\frac{1}{2}hQ_\alpha \psi_\alpha -h\mathcal{I}[\sigma](1/2)QA_{\beta \beta}+h\mathcal{I}[\mathcal{I}[\sigma]](1/2)P_\alpha A_{\beta \beta, \alpha}
\\
+h\mathcal{I}[\sigma \zeta](1/2)PB_{\alpha \alpha}+h^2(\frac{1}{2}\alpha -\mathcal{I}[\mathcal{I}[\sigma \zeta]](1/2))Q_\alpha B_{\beta \beta, \alpha}
\\
+h\mathcal{I} \Bigl[ \frac{\langle k\rangle k}{\langle k^2/\mu \rangle \mu}-1\Bigr] (1/2)Q_\alpha \bar{\varphi}_\alpha \Bigr\} \dd{a}\dd{t},
\end{multline*} 
where
\begin{equation*}
\begin{split}
P_\alpha =\tau_\alpha |_{\zeta=1/2} +\tau_\alpha |_{\zeta=-1/2}, \quad P=\tau |_{\zeta=1/2} +\tau |_{\zeta=-1/2},
\\
Q_\alpha =\tau_\alpha |_{\zeta=1/2} -\tau_\alpha |_{\zeta=-1/2}, \quad Q=\tau |_{\zeta=1/2} -\tau |_{\zeta=-1/2}.
\end{split}
\end{equation*}
The fifth and seventh terms can be transformed by integration by parts. In terms of the new unknown functions $\bar{u}$ and $\psi_\alpha$ the work becomes finally
\begin{multline}\label{eq:19}
\mathcal{A}=\int_{t_0}^{t_1}\int_{\Omega} \Bigl\{ P_\alpha u_\alpha +P \bar{u} +\frac{1}{2}hQ_\alpha \psi_\alpha -h\mathcal{I}[\sigma](1/2)QA_{\beta \beta}-h\mathcal{I}[\mathcal{I}[\sigma]](1/2)P_{\alpha ,\alpha}A_{\beta \beta}
\\
+h\mathcal{I}[\sigma \zeta](1/2)P B_{\beta \beta}-h^2\alpha PB_{\beta \beta}-h^2(\frac{1}{2}\alpha -\mathcal{I}[\mathcal{I}[\sigma \zeta]](1/2))Q_{\alpha ,\alpha}B_{\beta \beta}
\\
+h\mathcal{I} \Bigl[ \frac{\langle k\rangle k}{\langle k^2/\mu \rangle \mu}-1\Bigr] (1/2)Q_\alpha (\psi_\alpha +\bar{u}_{,\alpha })\Bigr\} \dd{a}\dd{t}.
\end{multline}

It is now appropriate to make a remark about the energy of the boundary layer. Unfortunately, this energy is currently unknown, and the neglect of the null-Lagrangian in the shortwave extrapolation makes its determination even more difficult. As a result, the asymptotically exact boundary conditions for the FSDT cannot be established. However, numerous numerical simulations for homogeneous plates show that the natural boundary conditions give sufficiently accurate results \citep{le1999vibrations}, so it is reasonable to expect the same for the FG-plates.

\section{Two-dimensional FSDT}

We formulate now the variational principle of the 2-D FSDT of free vibration of the FG-plate: the true displacements and rotation angles of the freely vibrating FG-plate change in space and time in such a way that the 2-D average action functional 
\begin{equation*}
J[u_\alpha ,\bar{u},\psi _\alpha ]=\int_{t_0}^{t_1}\int_{\Omega }[\Theta (\dot{u}_\alpha, \dot{\bar{u}},\dot{\psi }_\alpha)-\Phi (A_{\alpha \beta },B_{\alpha \beta },\psi_{\alpha }+\bar{u}_{,\alpha })]\dd{a} \dd{t},
\end{equation*} 
becomes stationary among all continuously differentiable functions $u_\alpha (x_\alpha ,t)$, $\bar{u}(x_\alpha ,t)$, and $\psi _\alpha(x_\alpha ,t)$. We assume that these functions are fixed at the initial and end time. If the edge of the plate is free, then no constraints are imposed on these functions at the boundary. The stored and kinetic energy densities are taken from \eqref{eq:17} and \eqref{eq:18}, respectively, with $A_{\alpha \beta}=u_{(\alpha ,\beta )}$ and $B_{\alpha \beta}=-\psi_{(\alpha ,\beta )}$. The standard calculus of variation shows that the stationarity condition $\delta J=0$ implies the following two-dimensional equations
\begin{equation}\label{eq:20}
\begin{split}
&\langle \rho \rangle h\ddot{u}_\alpha -\rho ^* h^3 \ddot{u}_{\beta,\beta \alpha }=n_{\alpha \beta ,\beta},
\\
&\langle \rho \rangle h\ddot{\bar{u}}=q_{\alpha ,\alpha },
\\
&\langle \rho \zeta^2+2\rho \alpha \rangle h^3 \ddot{\psi}_\alpha =-m_{\alpha \beta ,\beta}-q_\alpha.
\end{split}
\end{equation}
The 2-D tensors $n_{\alpha \beta}$, $q_\alpha$, and $m_{\alpha \beta}$ are given by the constitutive equations 
\begin{equation}\label{eq:21}
\begin{split}
&n_{\alpha \beta }=\pdv{\Phi}{A_{\alpha \beta}}=2\langle \mu \sigma \rangle h A_{\gamma \gamma} \delta_{\alpha \beta}+2\langle \mu \rangle h A_{\alpha \beta},
\\
&q_{\alpha }=\pdv{\Phi}{\bar{\varphi}_{\alpha }}=\mu ^* h(\psi_\alpha +\bar{u}_{,\alpha }),
\\
&m_{\alpha \beta }=\pdv{\Phi}{B_{\alpha \beta}}=2\langle \mu \sigma \zeta^2 \rangle h^3 B_{\gamma \gamma} \delta_{\alpha \beta}+2\langle \mu \zeta^2 \rangle h^3 B_{\alpha \beta},
\end{split}
\end{equation}
with $\delta_{\alpha \beta}$ being the Kronecker delta. For the plate with the free edge, these equations are subjected to the natural boundary conditions
\begin{equation}\label{eq:22} 
n_{\alpha \beta }\nu _{\beta }=0, \quad q_\alpha \nu_\alpha =0, \quad
m_{\alpha \beta }\nu _\beta =0,
\end{equation}
with $\nu _\alpha $ being the components of the unit vector in the $(x_1,x_2)$-plane normal to the curve $\partial \Omega$. For the plate with the clamped or simply supported edge, the natural boundary conditions \eqref{eq:22} must be replaced by the corresponding kinematical boundary conditions. If there are tractions acting on the faces of the plate so that we have forced vibration, then the work $\mathcal{A}$ should be added to the action functional. This leads to additional terms on the right-hand sides of Eqs.~\eqref{eq:20} that can easily be obtained by varying the work \eqref{eq:19} (see, e.g., \citet{berdichevsky2009variational}). Note that the system of equations \eqref{eq:20}, \eqref{eq:21} and boundary conditions \eqref{eq:22} reduces to those obtained by \citet{reissner1945the} and \citet{berdichevsky1979variational} for homogeneous plates with constant $\rho$, $\lambda$, and $\mu$.

To complete the 2-D FSDT of FG-plates we should also indicate the method of restoring the 3-D stress and strain states by means of the 2-D fields. To do this, the strain tensor field $\boldsymbol{\varepsilon}(x_\alpha ,\zeta,t)$ should be found from \eqref{eq:8}. Using the asymptotic formulas \eqref{eq:9}, \eqref{eq:10}, and \eqref{eq:12}, we obtain
\begin{equation}\label{eq:23}
\begin{split}
&\varepsilon _{\alpha \beta }=A_{\alpha \beta }- h\zeta B_{\alpha \beta },
\quad
2\varepsilon _{\alpha 3}=  \frac{\langle k\rangle k}{\langle k^2/\mu \rangle \mu} \bar{\varphi}_\alpha ,
\\
&\varepsilon _{33}=-\sigma A_{\alpha \alpha }+ \sigma \zeta h B_{\alpha \alpha }.
\end{split}
\end{equation}
In contrast to $\varepsilon_{\alpha \beta}$, the shear strains $\varepsilon _{\alpha 3}$ and the normal strain $\varepsilon_{33}$ turn out to be nonlinear functions of $\zeta$ due to the dependence of $\mu$ and $\sigma$ on $\zeta$. The stress tensor field $\boldsymbol{\sigma }(x_\alpha ,\zeta,t)$ is then determined by the 3-D constitutive equation (Hooke's law). While doing so, it is convenient to use the decomposition \eqref{eq:4} for the stored energy density. Within this approximation we find 
\begin{equation}\label{eq:24} 
\begin{split}
&\sigma_{\alpha \beta }=2\mu \sigma A_{\gamma \gamma}\delta _{\alpha \beta} +2\mu A_{\alpha \beta }-2\mu \sigma \zeta hB_{\gamma \gamma}\delta _{\alpha \beta} -2\mu \zeta hB_{\alpha \beta },
\\
&\sigma_{\alpha 3}=\frac{\langle k\rangle k}{\langle k^2/\mu \rangle }\bar{\varphi}_\alpha ,\quad \sigma_{33}=0,
\end{split}
\end{equation}
We see that, due to the dependence of $\mu$ and $\sigma$ on $\zeta$, $\sigma_{\alpha \beta}$ and $\sigma_{\alpha 3}$ are not polynomials of $\zeta$, as in the case of homogeneous plates. However, integrating these stresses over the thickness of the plate, we get
\begin{equation}
\label{eq:25}
\langle \sigma_{\alpha \beta }\rangle =\frac{n_{\alpha \beta }}{h}, \quad \langle \sigma_{\alpha \beta } \zeta \rangle =-\frac{m_{\alpha \beta }}{h^2}, \quad \langle \sigma_{\alpha 3}\rangle =\frac{q_\alpha }{h}.
\end{equation}
Eqs.~\eqref{eq:25} give a clear physical interpretation of these integral characteristics: $n_{\alpha \beta}$ are the stress resultants, $m_{\alpha \beta}$ are the bending moments, and $q_\alpha$ are the shear forces. For homogeneous plates the distributions \eqref{eq:23} and \eqref{eq:24} reduce to those obtained by \citet{berdichevsky1979variational}, but differ from those by \citet{reissner1945the}. Concerning the integral characteristics \eqref{eq:25} one can see the full agreement with both cited works for homogeneous plates.

With the purpose of applying the theory to special FG-plates let us introduce the short-hand notations
\begin{equation}
\label{eq:26}
\begin{split}
&\rho_1=\langle \rho \rangle,\quad \rho_2=\rho ^*, \quad \rho_3=\langle \rho \zeta^2+2\rho \alpha \rangle ,\quad \mu_1=\langle \mu \rangle,
\\
&\mu_2=\langle \mu \sigma \rangle,\quad \mu_3=\langle \mu \zeta^2\rangle, \quad \mu_4=\langle \mu \sigma \zeta^2\rangle,\quad \mu_5=\mu^*=\frac{\langle k\rangle ^2}{\langle k^2/\mu \rangle }.
\end{split}
\end{equation}
Thus, together with $\alpha$ the total number of coefficients of the 2-D FSDT is nine. Using these short-hand notations we present the system \eqref{eq:20} and \eqref{eq:21} in the form
\begin{equation}\label{eq:27}
\begin{split}
&\rho _1 h\ddot{u}_\alpha - \rho_2 h^3 \ddot{u}_{\beta ,\beta \alpha }=n_{\alpha \beta ,\beta}
\\
&\rho _1 h\ddot{\bar{u}}=q_{\alpha ,\alpha },
\\
&\rho _3 h^3 \ddot{\psi}_\alpha =-m_{\alpha \beta ,\beta}-q_\alpha,
\end{split}
\end{equation}
and 
\begin{equation}\label{eq:28}
\begin{split}
&n_{\alpha \beta }=2\mu _2 h A_{\gamma \gamma} \delta_{\alpha \beta}+2\mu _1 h A_{\alpha \beta}
\\
&q_{\alpha }=\mu_5 h (\psi_\alpha +\bar{u}_{,\alpha }),
\\
&m_{\alpha \beta }=2\mu _4 h^3 B_{\gamma \gamma} \delta_{\alpha \beta}+ 2\mu _3 h^3 B_{\alpha \beta}.
\end{split}
\end{equation}

\section{Wave propagation in sandwich plates}

\begin{figure}[htb]
	\centering
	\includegraphics[width=8cm]{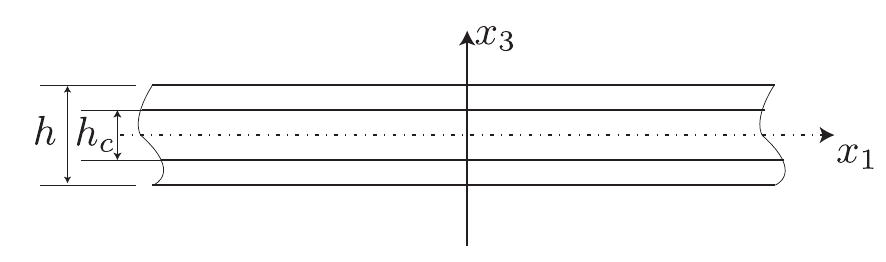}
	\caption{Cross section of a sandwich plate}
	\label{fig:2}
\end{figure}

Since the sandwich plate is a special case of FG-plates that admits an analytical solution to the 3-D problem of wave propagation \citep{lee1979harmonic} (see also \citep{kaplunov2017}), we can verify the asymptotic accuracy of our theory by applying it to this test problem. Consider an infinite sandwich plate, shown in Fig.~\ref{fig:2}, whose elastic moduli are given by
\begin{equation*}
\lambda(\zeta)=\begin{cases}
   \lambda_c,   & \text{$|\zeta|<b/2$}, \\
   \lambda_s,   & \text{$b/2<|\zeta|<1/2$}
\end{cases}, \quad
\mu(\zeta)=\begin{cases}
   \mu_c,   & \text{$|\zeta|<b/2$}, \\
   \mu_s,   & \text{$b/2<|\zeta|<1/2$}.
\end{cases}
\end{equation*}
The mass density and the coefficient $\sigma=\lambda/(\lambda+2\mu)$ have similar distributions
\begin{equation*}
\rho (\zeta)=\begin{cases}
   \rho_c,   & \text{$|\zeta|<b/2$}, \\
   \rho_s,   & \text{$b/2<|\zeta|<1/2$}
\end{cases}, \quad
\sigma(\zeta)=\begin{cases}
   \sigma_c,   & \text{$|\zeta|<b/2$}, \\
   \sigma_s,   & \text{$b/2<|\zeta|<1/2$}.
\end{cases}
\end{equation*}
Here $b=h_c/h$, with $h_c$ being the height of the core layer. Based on these distributions we can now compute all coefficients of the FSDT. Long, but otherwise elementary calculations according to \eqref{eq:26} give
\begin{equation*}
\begin{split}
&\rho_1=b(\rho_c-\rho_s)+\rho_s,
\\
&\rho_2=\frac{1}{12\sigma_s}[b^3\sigma_c^2(\rho_c\sigma_s-\rho_s\sigma_s)+\rho_s(b(\sigma_c-\sigma_s)+\sigma_s)^3]-\frac{\rho_1}{12(\mu_1+\mu_2)}(1-b)b
\\
&\times [b^2(\sigma_c-\sigma_s)-\sigma_s+b(2\sigma_s-3\sigma_c)](\mu_c \sigma_c-\mu_s \sigma_s+\mu_c-\mu_s),
\\
&\rho_3=\frac{1}{12}[b^3(\rho_c-\rho_s)+\rho_s]+2\alpha [b(\rho_c-\rho_s)+\rho_s],
\\
&\mu_1=b(\mu_c-\mu_s)+\mu_s,
\\
&\mu_2=b(\mu_c\sigma_c-\mu_s\sigma_s)+\mu_s\sigma_s,
\\
&\mu_3=\frac{1}{12}[b^3(\mu_c-\mu_s)+\mu_s],
\\
&\mu_4=\frac{1}{12}[b^3(\mu_c\sigma_c-\mu_s\sigma_s)+\mu_s\sigma_s],
\\
&\mu_5=\frac{20\mu_c\mu_s}{3(1-b)^3(8+9b+3b^2)\mu_c+3b(15-10b^2+3b^4)\mu_s}
\\
&\alpha=\frac{1}{120}[-5b^3(\sigma_c-\sigma_s)+3b^5(\sigma_c-\sigma_s)-2\sigma_c].
\end{split}
\end{equation*}
If $b=1$ and the $\rho$, $\mu$, and $\sigma$ values of the core and skin layers are the same (homogeneous plate), it is easy to verify that these coefficients reduce to those in \citep{berdichevsky1979variational}.

For sandwich plates (and generally for FG-plates with the even distributions of $\rho$, $\mu$, and $\sigma$), the system of equations \eqref{eq:25} and \eqref{eq:26} decomposes into the equations of the longitudinal and flexural waves. Let us first consider the equations \eqref{eq:25}$_{1}$ and \eqref{eq:26}$_{1}$, which describe the longitudinal waves. 
 In terms of $u_\alpha$ the equation of motion becomes
\begin{equation}\label{eq:29}
\rho _1 h\ddot{u}_\alpha - \rho_2 h^3 \ddot{u}_{\beta ,\beta \alpha }=h[(2\mu_2+\mu_1)u_{\beta ,\beta \alpha}+\mu_1\laplacian u_\alpha ],
\end{equation}
where $\laplacian$ is 2-D Laplace operator. We introduce the dimensionless time and coordinates
\begin{equation}
\label{eq:30}
\tau =\frac{t}{h}\sqrt{\frac{\mu _c}{\rho _c}},
\quad \zeta _\alpha =\frac{x_\alpha }{h}
\end{equation}
and rewrite Eq.~\eqref{eq:29} in the form
\begin{equation}\label{eq:31}
r_1 u_{\alpha |\tau \tau}- r_2 u_{\beta |\beta \alpha \tau \tau}=(2m_2+m_1)u_{\beta |\beta \alpha}+m_1\laplacian u_\alpha .
\end{equation}
Here and in the sequel the vertical bar preceding indices denotes the derivatives with respect to the corresponding dimensionless variables and, for simplicity, we use the same $\nabla ^2$ to denote the ``dimensionless'' Laplace operator. The dimensionless coefficients are
\begin{equation*}
r_i=\frac{\rho_i}{\rho_c},\quad i=1,2,3;
\quad
m_i=\frac{\mu_i}{\mu_c},\quad i=1,2,3,4,5.
\end{equation*}
Let us seek the solutions of \eqref{eq:31} in form of the harmonic plane waves propagating in the direction $\zeta _1$, $u_\alpha =a_\alpha e^{i(\kappa \zeta _1-\vartheta \tau )}$. It is easy to show that there are two possible types of waves corresponding to
\begin{align*}
&u_2=a_2e^{i(\kappa \zeta _1-\vartheta \tau )}, \quad u_1=0 \quad \text{SS-wave}
\\
&u_1=a_1e^{i(\kappa \zeta _1-\vartheta \tau )}, \quad u_2=0 \quad \text{L-wave}.
\end{align*}

\begin{figure}[htb]
	\centering
	\includegraphics[width=7cm]{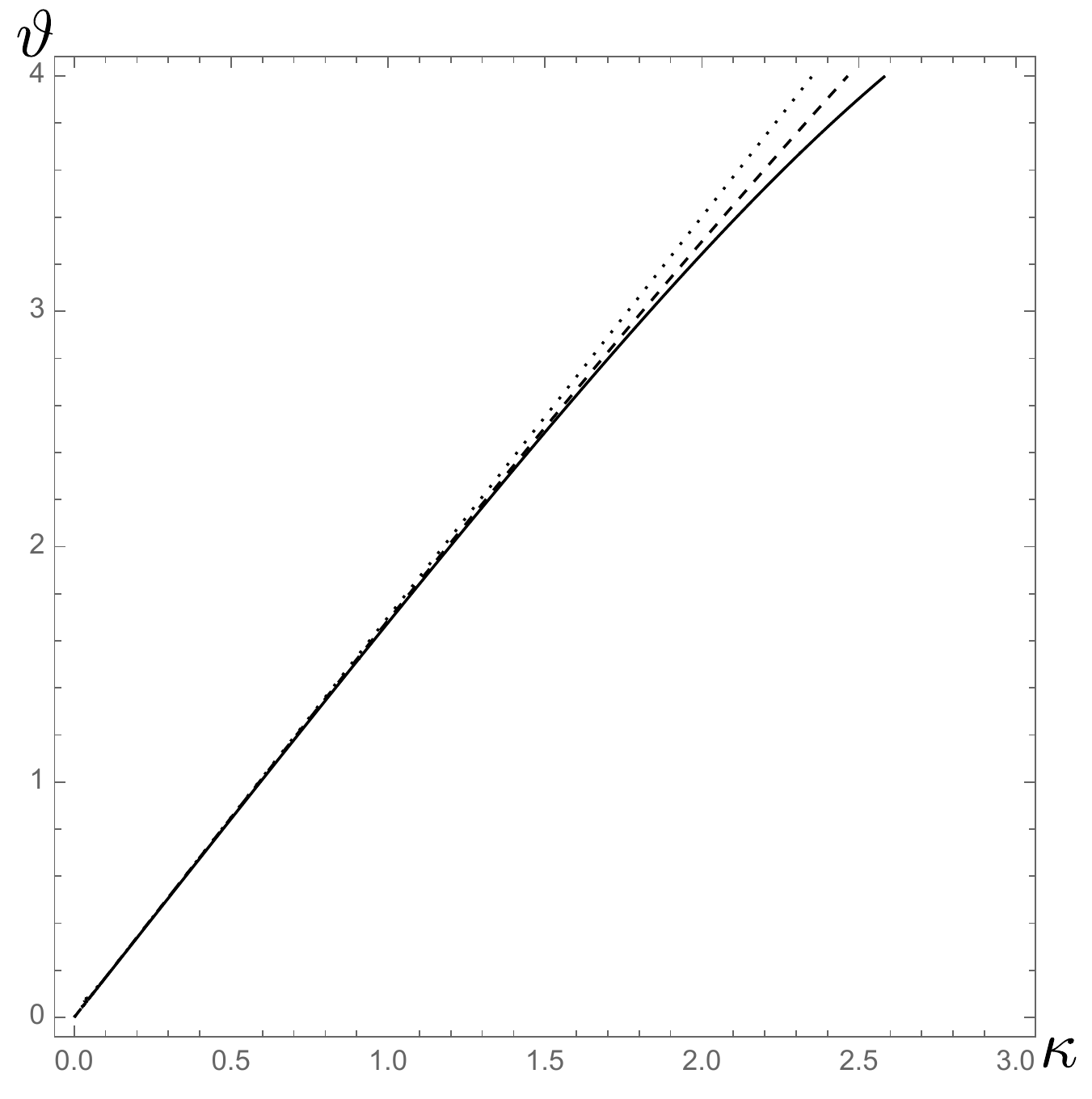}
	\caption{Dispersion curve of longitudinal waves in sandwich plate: (i) Bold line: theory of elasticity, (ii) Dashed line: FSDT for sandwich plate, (iii) Dotted line: Classical lamination theory for sandwich plate. The chosen parameters: $b=0.8$, $m_s=0.8$, $r_s=0.8$, $\nu_c=0.3$, $\nu_s=0.35$.}
	\label{fig:3}
\end{figure}

For the SS-waves (symmetric shear waves) the dispersion relation reads
\begin{equation*}
r_1\vartheta^2=m_1 \kappa^2
\end{equation*}
that coincides with the exact dispersion relation of the theory of elasticity \citep{lee1979harmonic}. For L-waves we have from \eqref{eq:31}
\begin{equation}
\label{eq:32}
r_1\vartheta^2+r_2\vartheta^2\kappa^2=2(m_1+m_2)\kappa^2.
\end{equation}
For homogenous plates this equation becomes
\begin{displaymath}
\vartheta^2+\frac{\sigma^2}{12}\vartheta^2\kappa^2=2(\sigma+1)\kappa^2.
\end{displaymath}
Bringing the second correction term on the left-hand side to the right-hand side, and replacing $\vartheta^2$ in it by $2(\sigma+1)\kappa^2$, which is allowed in this approximation, we obtain the asymptotic formula for the dispersion relation of the refined plate theory
\begin{displaymath}
\vartheta^2=2(\sigma+1)\kappa^2-\frac{\sigma^2(\sigma+1)}{6}\kappa^4=\frac{2}{1-\nu}\kappa^2-\frac{\nu^2}{6(1-\nu)^3}\kappa^4.
\end{displaymath}
This formula can also be derived from the Rayleigh-Lamb dispersion relation of elasticity theory \citep{le1999vibrations}, which means that the FSDT in this particular case is asymptotically accurate up to terms of order $h^2/l^2$. As for the dispersion relation \eqref{eq:32} for sandwich plates, we find that the first and last terms in this equation agree with the asymptotic formula for the dispersion relation from elasticity theory, which is the vanishing determinant of the $6\times 6$ matrix ($V$-matrix) \citep{lee1979harmonic}. It is also possible to check the validity of the second correction term in equation ~\eqref{eq:32} by expanding the determinant after the Taylor series expansion of the sine and cosine functions of the elements of this matrix. Since this is quite tedious and difficult to find the main asymptotic terms in this huge expression, we decide to check this by numerical simulation of the dispersion curves. Looking at Lee-Chang dispersion relation and Eq.~\eqref{eq:32}, we see that there are five parameters determining each dispersion curve:
\begin{displaymath}
b,\quad \sigma_c=\frac{\nu_c}{1-\nu_c},\quad \sigma_s=\frac{\nu_s}{1-\nu_s}, \quad r_s=\frac{\rho_s}{\rho_c},\quad m_s=\frac{\mu_s}{\mu_c}.
\end{displaymath}
Fig.~\ref{fig:3} shows a representative result of the numerical simulations of the dispersion curves, where the bold line corresponds to the dispersion curve obtained from the elasticity theory and the dashed and dotted lines represent the dispersion curves of the FSDT and the classical lamination theory for sandwich plates, respectively. The parameters chosen for these numerical simulations are: $b=0.8$, $\nu_c=0.3$, $\nu_s=0.35$, $r_s=0.8$, $m_s=0.8$. As we can see, the FSDT agrees much better with the dispersion curve of the elasticity theory than the classical lamination theory for sandwich plates. The excellent match even extends to the shortwave range, where $\kappa>1$. The variation of the parameters does not change this feature.

Let us now consider equations \eqref{eq:27}$_{2,3}$ and \eqref{eq:28}$_{2,3}$ describing the flexural waves. We look for the harmonic plane waves propagating in $x_1$-direction such that $\psi_1$ and $\bar{u}$ are functions of $x_1$ and $t$, while $\psi_2=0$. Then the system of equations \eqref{eq:27}$_{2,3}$ and \eqref{eq:28}$_{2,3}$ reduces to
\begin{equation}
\label{eq:33}
\begin{split}
&\rho _1 h\ddot{\bar{u}}=\mu_5 h (\psi_{1,1} +\bar{u}_{,11}),
\\
&\rho _3 h^3 \ddot{\psi}_1 =2(\mu_3+\mu_4)\psi_{1,11}-\mu_5 h (\psi_1 +\bar{u}_{,1}).
\end{split}
\end{equation}
We introduce the dimensionless variables $\tau $ and $\zeta _1$ in accordance with \eqref{eq:30} and $\bar{\psi}=h\psi $ and rewrite Eqs.~\eqref{eq:33} in the form
\begin{equation}
\label{eq:34}
\begin{split}
&r_1\bar{u}_{|\tau \tau}=m_5 (\bar{\psi}_{1|1} +\bar{u}_{|11}),
\\
&r_3 \bar{\psi}_{1|\tau \tau} =2(m_3+m_4)\bar{\psi}_{1|11}-m_5 (\bar{\psi}_1 +\bar{u}_{|1}).
\end{split}
\end{equation}
We seek solutions of these equations in form of the harmonic waves
\begin{equation*}
(\bar{u},\bar{\psi }_1)=(a,b)e^{i(\kappa \zeta _1-\vartheta \tau )}
\end{equation*}
and substitute it into \eqref{eq:34}. The condition of non-triviality of the solutions yields the dispersion relation
\begin{equation}\label{flex}
(m_5 \kappa ^2-r_1\vartheta ^2)[2(m_3+m_4)\kappa ^2 -r_3 \vartheta ^2
+m_5 ]-m_5^2 \kappa ^2=0.
\end{equation}
Expanding the left-hand side, we get
\begin{displaymath}
2m_5(m_3+m_4)\kappa^4-[m_5r_3+2(m_3+m_4)r_1]\kappa^2\vartheta^2-m_5r_1\vartheta^2+r_1 r_3\vartheta^4=0.
\end{displaymath}
Within the desired approximation the last term on the left-hand side of this equation can be neglected, while in the second correction term $\vartheta^2$ can be replaced by $2(m_3+m_4)\kappa^4/r_1$. This leads to the following asymptotic formula
\begin{equation}
\label{eq:35}
\vartheta^2=\frac{2(m_3+m_4)}{r_1}\kappa^4-2(m_3+m_4)\Bigl( \frac{r_3}{r_1}+2\frac{m_3+m_4}{m_5}\Bigr) \kappa^6.
\end{equation}
For homogenous plates this asymptotic formula becomes
\begin{equation*}
\vartheta^2=\frac{1}{6(1-\nu)}\kappa^4-\frac{17-7\nu}{360(1-\nu)^2}\kappa^6.
\end{equation*}
Note that this same formula can also be derived from the Rayleigh-Lamb dispersion relation of elasticity theory \citep{le1999vibrations}, which means that the FSDT for bending and flexural vibration of homogeneous plates is asymptotically exact up to terms of order $h^2/l^2$. 

\begin{figure}[htb]
	\centering
	\includegraphics[width=7cm]{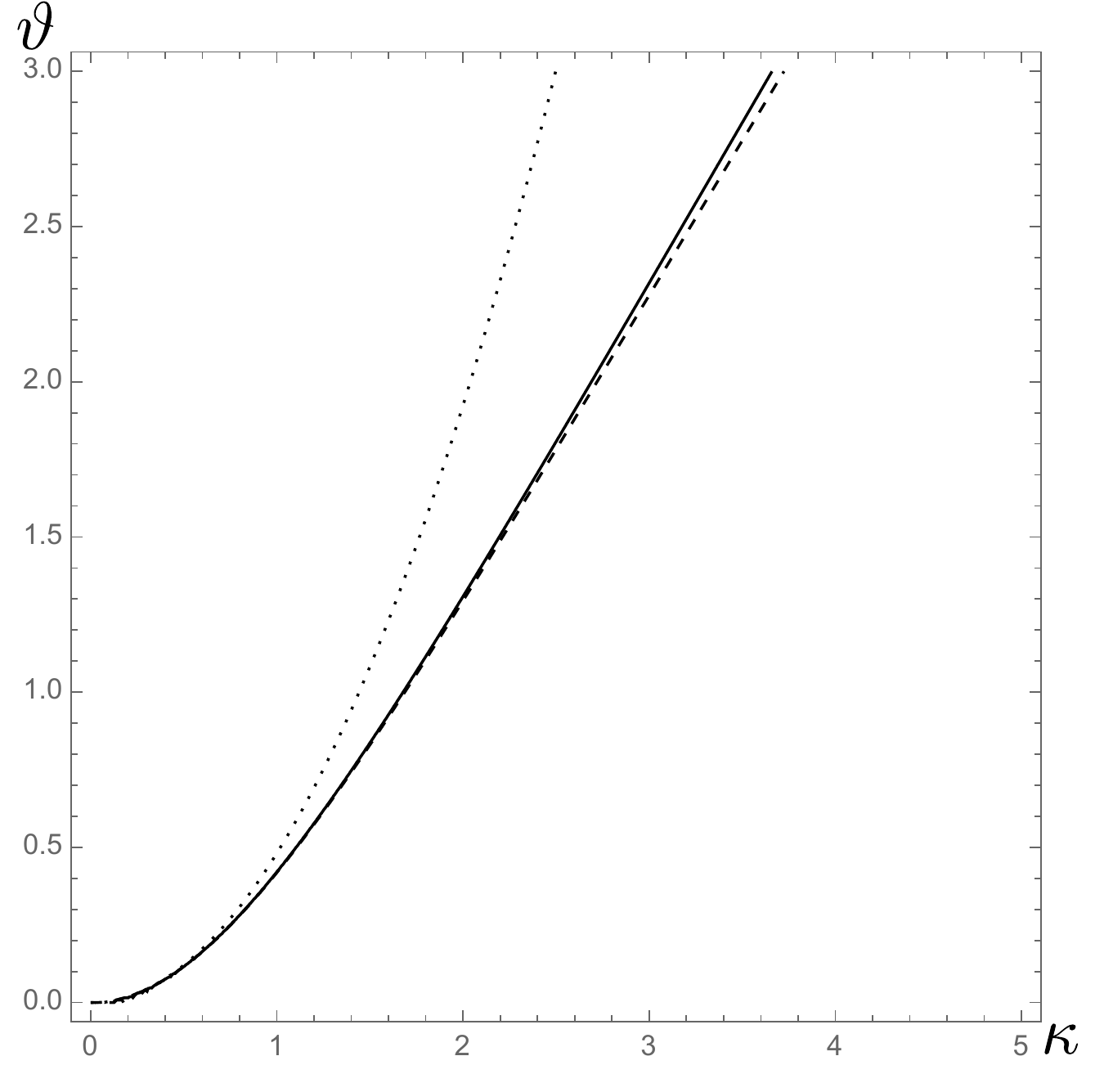}
	\caption{Dispersion curve of flexural waves in sandwich plate: (i) Bold line: theory of elasticity, (ii) Dashed line: FSDT for sandwich plate, (iii) Dotted line: Classical lamination theory for sandwich plate. The chosen parameters: $b=0.8$, $m_s=0.8$, $r_s=0.8$, $\nu_c=0.3$, $\nu_s=0.35$.}
	\label{fig:4}
\end{figure}

As for the dispersion relation \eqref{eq:35} for sandwich plates, we note that the first term on the right-hand side agree with the asymptotic formula for the dispersion relation from elasticity theory \citep{lee1979harmonic}. It is also possible to check the validity of the second correction term in Eq.~\eqref{eq:35} by expanding the determinant of the $6\times 6$ matrix ($W$-matrix) after the Taylor series expansion of the trigonometric functions of the elements of this matrix.\footnote{There is a misprint in \citep{lee1979harmonic} for $W$-matrix: in the element $w_{56}$ of this matrix $b_1$ should be replaced by $b_2$.} Similar considerations as in the previous case convinced us that the better way to check this is to do numerical simulations of the dispersion curves. Fig.~\ref{fig:4} shows a representative result of these numerical simulations, where the bold line corresponds to the dispersion curve obtained from the elasticity theory and the dashed and dotted lines represent the dispersion curves of the FSDT in accordance with \eqref{flex} and the classical lamination theory for sandwich plates. The parameters chosen for these numerical simulations remain the same as in the previous case. As we can see, the FSDT agrees much better with the dispersion curve of the elasticity theory than the classical lamination theory for sandwich plates. The excellent agreement between the exact curve and that of the FSDT is observed even in the short-wave range, up to $\kappa=3$. Again, the variation of the parameters does not change this feature. This verification confirms the asymptotic exactness of the FSDT up to order $h^2/l^2$ for sandwich plates, and to our conviction also for FG-plates. It is also interesting to note that, as observed by \citet{kuznetsov2021}, the dispersion curves of guided waves in plates with even (asymmetric layout) or odd (symmetric layout) number of alternating layers are significantly different and remain different even with a large number of alternating layers. Since such a plate can be considered as an FG-plate, the developed variational-asymptotic method leads to asymptotic accuracy of FSDT only for the symmetric layout, but not for the asymmetric one.

\section{Conclusion}
In this work, it has been demonstrated that the asymptotically exact FSDT of FG-plates whose mass density and elastic moduli are even functions of the transverse coordinate can be derived from the 3-D elasticity theory by the variational-asymptotic method. Comparison of the dispersion curves obtained from the elasticity theory and the FSDT for sandwich plates shows that the latter is accurate up to an order of magnitude $h^2/l^2$ compared to unity. The generalization to anisotropic elastic FG-shells, which brings further terms describing the cross effect between extension and bending, is straightforward. It is also not difficult to include the geometrically nonlinear effect when the displacements become large. The open question is how to derive the FSDT for FG-plates and shells whose mass density and elastic moduli are not even functions of the transverse coordinate. As discussed at the end of Section 4, the difference between symmetric and asymmetric layouts limitates our asymptotic analysis, with the likely consequence that there is no FSDT for FG-plates/shells with asymmetric layouts, and one must construct the FSDT that most closely approximates the asymptotically exact theory of the plate/shell \citep{sutyrin1997derivation,yu2005mathematical}. Another extremely important question is the numerical implementation of the refined plate theory that is free of shear-locking effect. This issue will be addressed in our next paper.

\end{document}